\documentstyle[12pt,epic,eepic]{article}

\def\SPACE{\vspace*{3ex}}

\newtheorem{theorem}{Theorem}
\newtheorem{lemma}{Lemma}
\newtheorem{corollary}{Corollary}

\def\QED{\mbox{\rule[0pt]{1.5ex}{1.5ex}}}

\def\endproof{\hspace*{\fill}~\QED\par\endtrivlist\unskip}
\def\keywords{\vspace{-.3em}
    \if@twocolumn
      \small\it Keywords\/\bf---$\!$%
    \else
      \begin{center}\small\bf Keywords\end{center}\quotation\small
    \fi}
\def\endkeywords{\vspace{0.6em}\par\if@twocolumn\else\endquotation\fi
    \normalsize\rm}
\def\appendix{\par
    \setcounter{section}{0}\setcounter{subsection}{0}
    \def\thesection{\Alph{section}} \section*{Appendix}
}
\renewcommand{\Re}{\mbox{\bf R}}

\newcommand{\Tr}{\mbox{\rm Tr\,}}
\renewcommand{\epsilon}{\varepsilon}
\renewcommand{\phi}{\varphi}
\def\Def{\stackrel{\rm def}=} % Def

\def\H{{\cal H}} % Hilbert space
\def\Hn{{\cal H}^{\otimes n}} % n tensor Hibert space
\def\S{{\cal S}} % the set of density operators
\def\B{{\cal B}} % the set of bounded operaters
\def\nrho{\rho^{\otimes n}} % n tensor \rho
\def\nsigma{\sigma^{\otimes n}} % n tensor \sigma
\def\Xl{X_{n,\lambda}} % X_{n,\lambda}
\def\X{{\cal X}} % the set of alphabets
\def\A{{\cal A}} % acceptance region

\def\argmax{\mathop{\rm argmax}}
\def\Im{\mbox{\rm Im}\,}

\def\tipsi{\tilde{\psi}} % tilde psi
\begin{document}

\title{
Strong Converse and Stein's Lemma \\
in the Quantum Hypothesis Testing
}

\author{Tomohiro Ogawa and Hiroshi Nagaoka\thanks{
The authors are with
the Graduate School of Information Systems,
University of Electro-Communications,
1--5--1 Chofugaoka, Chofu, Tokyo 182--8585, Japan.
(E-mail: ogawa@hn.is.uec.ac.jp, nagaoka@is.uec.ac.jp)
}
}

\date{}

\maketitle

\begin{abstract}
The hypothesis testing problem of two quantum states is treated.  We
show a new inequality between the error of the first kind and the second
kind, which complements the result of Hiai and Petz to establish the
quantum version of Stein's lemma.  The inequality is also used to show a
bound on the first kind error when the power exponent for the second
kind error exceeds the quantum relative entropy, and the bound yields
the strong converse in the quantum hypothesis testing.  Finally, we
discuss the relation between the bound and the power exponent derived by
Han and Kobayashi in the classical hypothesis testing.
\end{abstract}

\begin{keywords}
Quantum hypothesis testing, Stein's lemma, strong converse,
quantum relative entropy.
\end{keywords}

%%%%%%%%%%%%%%%%%%%%%%
\section{Introduction}
%%%%%%%%%%%%%%%%%%%%%%

Let $\H$ be a Hilbert space
which represents a physical system in interest.
We suppose $\dim\H<\infty$ for mathematical simplicity.
Let $\B(\H)$ be the set of linear operators on $\H$ and put
\begin{eqnarray*}
\S(\H) \Def \{ \rho\in\B(\H) \,|\, \rho=\rho^*\ge 0, \Tr\rho=1 \} ,
\end{eqnarray*}
which is the set of density operators on $\H$.

We treat the problem of hypothesis testing a null hypothesis
$\rho\in\S(\H)$ versus an alternative hypothesis $\sigma\in\S(\H)$.
Here, we assume $\Im \rho \subset \Im \sigma$.
To consider an asymptotic situation, suppose that either $\nrho \in
\S(\Hn)$ or $\nsigma \in \S(\Hn)$ is given.  The problem is to decide
which hypothesis is true, and the decision is given by a two-valued
quantum measurement $\{A_n, 1-A_n\}\, (A_n\in \B(\Hn),0\le A_n\le 1)$,
where $A_n$ corresponds to the acceptance of $\nrho$ and $1-A_n$
corresponds to the acceptance of $\nsigma$.  We call $A_n\in \B(\Hn)$
($0\le A_n\le 1$) a test in the sequel.

For a test $A_n$, define the error probability of the first kind and
the second kind by
\begin{eqnarray*}
 \alpha_n (A_n) &\Def& \Tr \nrho (1-A_n) , \\
 \beta_n (A_n) &\Def& \Tr \nsigma A_n ,
\end{eqnarray*}
respectively. We see that $\alpha_n (A_n)$ is the error probability of
the acceptance of $\nsigma$ when $\nrho$ is true and $\beta_n (A_n)$ is
the error probability of the converse situation.  Since we can not have
$\alpha_n (A_n)$ and $\beta_n(A_n)$ arbitrarily small simultaneously, we
will make $\beta_n (A_n)$ as small as possible under the constraint
$\alpha_n (A_n)\le\epsilon$.  In other words,
the problem is to examine the asymptotic behavior of the following
quantity:
\begin{eqnarray*}
\beta^*_n (\epsilon) \Def \min \{ \beta_n(A_n) \,|\,
A_n \in \B (\Hn),\, 0\le A_n \le I,\, \alpha_n(A_n) \le \epsilon \} .
\end{eqnarray*}

Concerning $\beta^*_n (\epsilon)$,
Hiai and Petz \cite{Hiai-Petz} showed
\begin{eqnarray}
\limsup_{n\rightarrow\infty} \frac{1}{n}\log \beta_n^*(\epsilon)
\le -D(\rho||\sigma) ,
\label{Hiai-Petz-sup}
\end{eqnarray}
and
\begin{eqnarray}
-\frac{1}{1-\epsilon} D(\rho||\sigma) \le
\liminf_{n\rightarrow\infty} \frac{1}{n}\log \beta_n^*(\epsilon) ,
\label{Hiai-Petz-inf}
\end{eqnarray}
where
\begin{eqnarray*}
D(\rho||\sigma) \Def \Tr \rho (\log \rho - \log \sigma) ,
\end{eqnarray*}
is the quantum relative entropy.
As for (\ref{Hiai-Petz-inf}),
they used the monotonicity of the quantum relative entropy
\cite{Lindblad-CP,Uhlmann} as follows:
\begin{eqnarray*}
&&D(\nrho||\nsigma) \\
&\ge& \alpha_n(A_n) \log\frac{\alpha_n(A_n)}{1-\beta_n(A_n)}
 + (1-\alpha_n(A_n)) \log\frac{1-\alpha_n(A_n)}{\beta_n(A_n)} \\
&=& -h(\alpha_n(A_n)) -\alpha_n(A_n)\log(1-\beta_n(A_n))
 - (1-\alpha_n(A_n)) \log \beta_n(A_n) \\
&\ge& -\log 2  - (1-\alpha_n(A_n)) \log \beta_n(A_n) ,
\end{eqnarray*}
where $h(x)$ is the binary entropy. Thus it holds that
\begin{eqnarray}
(1-\alpha_n(A_n))\frac{1}{n} \log \beta_n(A_n)
\ge -\frac{\log 2}{n} -D(\rho||\sigma) ,
\label{weak-converse}
\end{eqnarray}
which immediately yields (\ref{Hiai-Petz-inf}).
Note that (\ref{weak-converse}) also
leads the weak converse property, which means that
if $\beta_n(A_n)\le e^{-nr}\,(r>D(\rho||\sigma))$ then
$\alpha_n(A_n)$ does not go to zero as $n\rightarrow\infty$.

In this paper, we will show a fundamental inequality, which complements
(\ref{Hiai-Petz-sup}) by Hiai and Petz to show the quantum version of
Stein's lemma (see {\it e.g.} \cite{Blahut-text}, p.115).
We will also show a bound on $1-\alpha_n(A_n)$ under the
exponential-type constraint $\beta_n(A_n)\le e^{-nr}$.  The bound leads
to the strong converse property \cite{Blahut,Han-Kobayashi} in the
quantum hypothesis testing, i.e., if $\beta_n(A_n)\le
e^{-nr}\,(r>D(\rho||\sigma))$ then $\alpha_n(A_n)$ goes to one as
$n\rightarrow\infty$.  Finally, we discuss the relation with the result
of Han and Kobayashi \cite{Han-Kobayashi} in the classical hypothesis
testing.

%%%%%%%%%%%%%%%%%%%%%%%%%%%%%%%%%%%%%%%%%%%%%%%%%%%%%%%%
\section{A Fundamental Bound on the Error Probabilities}
%%%%%%%%%%%%%%%%%%%%%%%%%%%%%%%%%%%%%%%%%%%%%%%%%%%%%%%%

In this section, we show a fundamental inequality
between the error probabilities of the first kind and the second kind.

Let $\lambda$ be a real number and
\begin{eqnarray}
&& \nrho-e^{n\lambda}\nsigma = \sum_j \mu_{n,j} E_{n,j} ,
\label{spectral}
\end{eqnarray}
be the spectral decomposition. Define a test $\Xl$ by
\begin{eqnarray*}
\Xl \Def \sum_{j\in D_n} E_{n,j} ,
\end{eqnarray*}
where $D_n=\{j \,|\, \mu_{n,j} \ge 0\}$.
Then, the following lemma holds,
which corresponds to the quantum version of the Neyman-Pearson lemma
(see \cite{Helstrom}, p.108).
%%%%%%%%%%%%%%%%%%%%%%%%%%%%%%%%%%%%%%%%%%%%%%%%%%%%%%%%%%%%%%%%%%%%%%%%
\begin{lemma}
For any test $A_n$, we have
\begin{eqnarray}
\Tr (\nrho-e^{n\lambda}\nsigma)\Xl \ge
\Tr (\nrho-e^{n\lambda}\nsigma)A_n .
\label{Neyman-Pearson}
\end{eqnarray}
\end{lemma}
%%%%%%%%%%%%%%%%%%%%%%%%%%%%%%%%%%%%%%%%%%%%%%%%%%%%%%%%%%%%%%%%%%%%%%%%
\begin{proof}
\begin{eqnarray*}
\Tr (\nrho-e^{n\lambda}\nsigma)A_n
&=& \sum_j \mu_{n,j} \Tr E_{n,j} A_n \\
&\le& \sum_{j\in D_n} \mu_{n,j} \Tr E_{n,j} A_n \\
&\le& \sum_{j\in D_n} \mu_{n,j} \Tr E_{n,j} \\
&=& \Tr (\nrho-e^{n\lambda}\nsigma)\Xl .
\end{eqnarray*}
\end{proof}
%%%%%%%%%%%%%%%%%%%%%%%%%%%%%%%%%%%%%%%%%%%%%%%%%%%%%%%%%%%%%%%%%%%%%%%%

%%%%%%%%%%%%%%%%%%%%%%%%%%%%%%%%%%%%%%%%%%%%%%%%%%%%%%%%%%%%%%%%%%%%%%%%
\begin{theorem}
For any test $A_n$ and any $\lambda\in\Re$, we have
\begin{eqnarray}
1-\alpha_n(A_n) \le e^{-n\phi(\lambda)} + e^{n\lambda} \beta_n(A_n) ,
\label{fundamental}
\end{eqnarray}
where
\begin{eqnarray}
\phi(\lambda) &\Def&
\max_{0 \le s \le 1} \{ \lambda s - \psi(s) \} , \\
\psi(s) &\Def& \log \Tr \rho^{1+s} \sigma^{-s} .
\end{eqnarray}
\label{theorem:fundamental}
\end{theorem}
%%%%%%%%%%%%%%%%%%%%%%%%%%%%%%%%%%%%%%%%%%%%%%%%%%%%%%%%%%%%%%%%%%%%%%%%

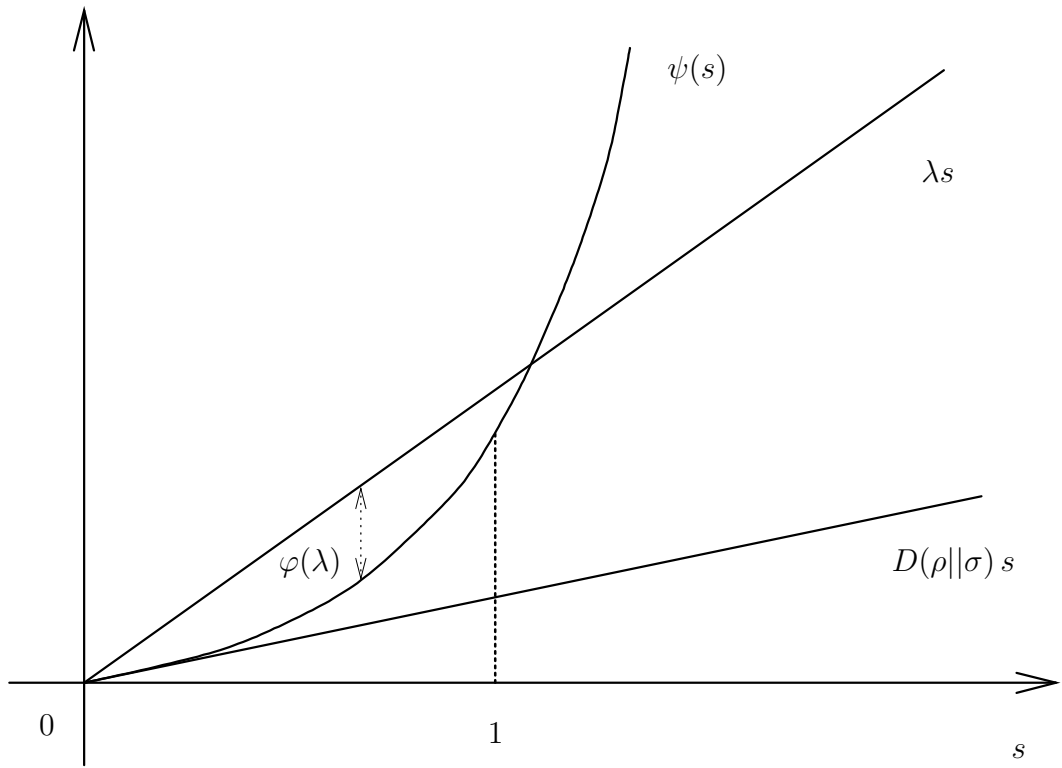
\begin{figure}[htbp]
\begin{center}
\setlength{\unitlength}{0.00087489in}
\begingroup\makeatletter\ifx\SetFigFont\undefined%
\gdef\SetFigFont#1#2#3#4#5{%
  \reset@font\fontsize{#1}{#2pt}%
  \fontfamily{#3}\fontseries{#4}\fontshape{#5}%
  \selectfont}%
\fi\endgroup%
{\renewcommand{\dashlinestretch}{30}
\begin{picture}(6885,4604)(0,-10)
\put(1642,1192){\makebox(0,0)[lb]{\smash{{{\SetFigFont{12}{14.4}{\familydefault}{\mddefault}{\updefault}$\phi(\lambda)$}}}}}
\thicklines
\path(472,517)(5872,1642)
\path(472,517)(5647,4207)
\path(22,517)(6322,517)
\path(6082.000,457.000)(6322.000,517.000)(6082.000,577.000)
\thinlines
\path(2107.000,1267.000)(2137.000,1147.000)(2167.000,1267.000)
\dottedline{45}(2137,1147)(2137,1687)
\path(2167.000,1567.000)(2137.000,1687.000)(2107.000,1567.000)
\thicklines
\dottedline{45}(2947,2017)(2947,517)
\path(472,517)(474,517)(477,518)
	(483,519)(493,522)(507,525)
	(525,529)(548,534)(575,540)
	(607,547)(642,555)(681,563)
	(722,573)(766,582)(811,593)
	(857,603)(903,614)(948,625)
	(993,635)(1036,646)(1077,656)
	(1117,666)(1155,676)(1191,686)
	(1225,695)(1257,704)(1288,714)
	(1317,723)(1345,732)(1372,742)
	(1401,753)(1429,764)(1457,775)
	(1484,786)(1511,797)(1538,809)
	(1565,821)(1591,833)(1617,845)
	(1643,856)(1669,869)(1694,881)
	(1719,893)(1745,905)(1770,918)
	(1795,930)(1820,943)(1845,956)
	(1870,969)(1895,982)(1920,996)
	(1945,1010)(1969,1024)(1994,1039)
	(2019,1054)(2043,1069)(2068,1085)
	(2092,1102)(2116,1119)(2140,1137)
	(2164,1155)(2187,1174)(2211,1193)
	(2235,1213)(2258,1233)(2282,1254)
	(2306,1275)(2329,1296)(2353,1318)
	(2377,1340)(2400,1362)(2424,1384)
	(2447,1406)(2471,1428)(2494,1451)
	(2517,1473)(2539,1495)(2561,1517)
	(2583,1538)(2605,1560)(2626,1582)
	(2646,1603)(2666,1624)(2685,1645)
	(2704,1666)(2722,1687)(2741,1709)
	(2759,1732)(2776,1754)(2792,1777)
	(2808,1799)(2823,1821)(2837,1843)
	(2852,1865)(2865,1886)(2878,1908)
	(2891,1930)(2904,1952)(2917,1973)
	(2929,1995)(2942,2017)(2954,2039)
	(2966,2061)(2979,2083)(2991,2105)
	(3004,2128)(3017,2151)(3030,2175)
	(3043,2198)(3056,2223)(3069,2247)
	(3082,2272)(3094,2295)(3106,2319)
	(3118,2343)(3130,2368)(3141,2393)
	(3153,2418)(3165,2444)(3176,2469)
	(3188,2495)(3200,2522)(3211,2548)
	(3223,2574)(3234,2601)(3246,2628)
	(3257,2655)(3269,2681)(3280,2708)
	(3291,2735)(3303,2761)(3314,2787)
	(3325,2814)(3335,2840)(3346,2865)
	(3357,2891)(3367,2917)(3377,2942)
	(3387,2967)(3397,2992)(3407,3019)
	(3417,3045)(3427,3072)(3437,3098)
	(3446,3125)(3455,3151)(3465,3177)
	(3474,3203)(3483,3229)(3492,3254)
	(3500,3280)(3509,3306)(3518,3332)
	(3527,3357)(3535,3383)(3544,3409)
	(3552,3434)(3560,3460)(3569,3486)
	(3577,3512)(3585,3537)(3592,3563)
	(3600,3589)(3608,3615)(3615,3641)
	(3622,3667)(3629,3693)(3635,3720)
	(3642,3747)(3649,3775)(3655,3805)
	(3662,3837)(3669,3870)(3676,3904)
	(3683,3941)(3690,3978)(3697,4017)
	(3705,4056)(3712,4094)(3719,4133)
	(3726,4169)(3732,4203)(3738,4234)
	(3743,4262)(3747,4285)(3750,4304)
	(3753,4319)(3755,4330)(3756,4337)
	(3757,4340)(3757,4342)
\put(6052,67){\makebox(0,0)[lb]{\smash{{{\SetFigFont{12}{14.4}{\familydefault}{\mddefault}{\updefault}$s$}}}}}
\put(5512,3532){\makebox(0,0)[lb]{\smash{{{\SetFigFont{12}{14.4}{\familydefault}{\mddefault}{\updefault}$\lambda s$}}}}}
\put(5332,1192){\makebox(0,0)[lb]{\smash{{{\SetFigFont{12}{14.4}{\familydefault}{\mddefault}{\updefault}$D(\rho||\sigma)\,s$}}}}}
\put(2902,157){\makebox(0,0)[lb]{\smash{{{\SetFigFont{12}{14.4}{\familydefault}{\mddefault}{\updefault}$1$}}}}}
\put(202,202){\makebox(0,0)[lb]{\smash{{{\SetFigFont{12}{14.4}{\familydefault}{\mddefault}{\updefault}$0$}}}}}
\put(3982,4162){\makebox(0,0)[lb]{\smash{{{\SetFigFont{12}{14.4}{\familydefault}{\mddefault}{\updefault}$\psi(s)$}}}}}
\path(472,22)(472,4567)
\path(532.000,4327.000)(472.000,4567.000)(412.000,4327.000)
\end{picture}
}
\end{center}
\caption{The graph of $\psi(s)$}
\label{fig:psi}
\end{figure}

Here, note that $\phi(\lambda)$ is the Legendre transformation
of a convex function $\psi(s)$
(see Fig. \ref{fig:psi} and \ref{fig:phi}).
Putting $A = \log\rho-\log\sigma-\psi'(s)$,
the convexity of $\psi(s)$ is verified as
\begin{eqnarray*}
\psi'(s) &=&
e^{-\psi(s)}\, \Tr \rho^{1+s}\sigma^{-s}(\log\rho-\log\sigma) , \\
\psi''(s) &=& e^{-\psi(s)}\, \Tr \rho^{1+s} A \sigma^{-s} A \\
&=& e^{-\psi(s)}\, \Tr
\left( \rho^{\frac{1+s}{2}} A \sigma^{-\frac{s}{2}} \right)
\left( \rho^{\frac{1+s}{2}} A \sigma^{-\frac{s}{2}} \right)^* \\
&>& 0 .
\end{eqnarray*}
Observing that $\psi(0)=0$ and $\psi'(0)=D(\rho||\sigma)$,
we can see that if $\lambda>D(\rho||\sigma)$ then $\phi(\lambda)>0$.
It is important to note that for any $\lambda$ satisfying
$D(\rho||\sigma)\le\lambda\le\psi'(1)$, we have
\begin{eqnarray}
s^*=\argmax_{0\le s \le 1} \{ \lambda s -\psi(s) \}
\Longleftrightarrow \psi'(s^*)=\lambda .
\label{Legendre-trans}
\end{eqnarray}

\SPACE
\noindent\hspace{2em}{\it Proof of Theorem \ref{theorem:fundamental}: }
Define probability distributions $p_n=\{p_{n,j}\}$ and
$q_n=\{q_{n,j}\}$ by
\begin{eqnarray*}
p_{n,j} = \Tr \nrho E_{n,j},\quad q_{n,j} = \Tr \nsigma E_{n,j} .
\end{eqnarray*}
From (\ref{spectral}), we have
$\mu_{n,j} \Tr E_{n,j} = p_{n,j} - e^{n\lambda} q_{n,j}$,
and hence,
\begin{eqnarray*}
D_n=\{j \,|\, 0\le\forall s\le 1,\,
      e^{-n\lambda s} p_{n,j}^s q_{n,j}^{-s} \ge 1 \} .
\end{eqnarray*}
Thus, we obtain
\begin{eqnarray}
\Tr \nrho \Xl &=& \sum_{j\in D_n} \Tr \nrho E_{n,j}
\nonumber \\
&=& \sum_{j\in D_n} p_{n,j}
\nonumber \\
&\le& \sum_{j\in D_n} p_{n,j} \cdot e^{-n\lambda s} p_{n,j}^s q_{n,j}^{-s}
\nonumber \\
&\le& e^{-n\lambda s} \sum_{j} p_{n,j}^{1+s} q_{n,j}^{-s}
\nonumber \\
&\le& e^{-n\lambda s}\, \Tr (\nrho)^{1+s} (\nsigma)^{-s} ,
\nonumber
\end{eqnarray}
where we used the monotonicity of the quantum $f$-divergence \cite{Petz}
for an operator convex function $f(u)=u^{-s}\,(0 \le s \le 1)$
(see {\it e.g.} \cite{Bhatia}, p.123).
Therefore, we have
\begin{eqnarray*}
\Tr \nrho \Xl 
\le \exp \left[ -n \{ \lambda s - \psi(s) \} \right] ,
\end{eqnarray*}
and hence,
\begin{eqnarray*}
\Tr \nrho \Xl \le e^{-n\phi(\lambda)} ,
\end{eqnarray*}
by taking the maximum.
Now, from (\ref{Neyman-Pearson}), the theorem is proved as follows:
\begin{eqnarray*}
1-\alpha_n(A_n)
&=& \Tr \nrho A_n \\
&\le& \Tr (\nrho-e^{n\lambda}\nsigma)\Xl + e^{n\lambda} \Tr \nsigma A_n \\
&\le& \Tr \nrho \Xl + e^{n\lambda} \Tr \nsigma A_n \\
&\le& e^{-n\phi(\lambda)} + e^{n\lambda} \beta_n(A_n) .
\end{eqnarray*}
\endproof

\section{The Quantum Stein's Lemma}

%%%%%%%%%%%%%%%%%%%%%%%%%%%%%%%%%%%%%%%%%%%%%%%%%%%%%%%%%%%%%%%%%%%%%%%%
\begin{theorem}
For any $0 \le \epsilon <1$, it holds that
\begin{eqnarray}
\lim_{n\rightarrow\infty} \frac{1}{n}\log \beta_n^*(\epsilon)
= -D(\rho||\sigma) .
\end{eqnarray}
\label{theorem:Stein}
\end{theorem}
%%%%%%%%%%%%%%%%%%%%%%%%%%%%%%%%%%%%%%%%%%%%%%%%%%%%%%%%%%%%%%%%%%%%%%%%
\begin{proof}
From (\ref{Hiai-Petz-sup}) by Hiai and Petz, we have only to show that
\begin{eqnarray}
\liminf_{n\rightarrow\infty} \frac{1}{n}\log \beta_n^*(\epsilon)
\ge -D(\rho||\sigma) .
\label{Stein-liminf}
\end{eqnarray}
Let $A_n$ be an arbitrary test which satisfies
$\alpha_n(A_n)\le\epsilon$.
From (\ref{fundamental}), we have
\begin{eqnarray*}
1-\epsilon \le 1-\alpha_n(A_n) \le 
e^{-n\phi(\lambda)} + e^{n\lambda} \beta_n(A_n) ,
\end{eqnarray*}
and hence,
\begin{eqnarray*}
\beta_n(A_n) \ge e^{-n\lambda} (1-\epsilon-e^{-n\phi(\lambda)}) .
\end{eqnarray*}
By taking the minimum, we obtain
\begin{eqnarray}
\beta_n^*(\epsilon) \ge e^{-n\lambda} (1-\epsilon-e^{-n\phi(\lambda)}) .
\label{beta-bound}
\end{eqnarray}
Now, let $\lambda=D(\rho||\sigma)+\delta\ (\delta>0)$, then
$\phi(\lambda)>0$ and $1-\epsilon-e^{-n\phi(\lambda)}>0$ for
sufficiently large $n$. Thus, (\ref{beta-bound}) yields
\begin{eqnarray*}
\frac{1}{n}\log \beta_n^*(\epsilon)
\ge -\lambda + \frac{1}{n}\log(1-\epsilon-e^{-n\phi(\lambda)}) ,
\end{eqnarray*}
and hence,
\begin{eqnarray*}
\liminf_{n\rightarrow\infty} \frac{1}{n}\log \beta_n^*(\epsilon)
\ge -D(\rho||\sigma) -\delta .
\end{eqnarray*}
Since $\delta>0$ is arbitrary,
the theorem has been proved.
\end{proof}
%%%%%%%%%%%%%%%%%%%%%%%%%%%%%%%%%%%%%%%%%%%%%%%%%%%%%%%%%%%%%%%%%%%%%%%%

\section{Strong Converse}

\begin{figure}[htbp]
\begin{center}
\setlength{\unitlength}{0.00087489in}
\begingroup\makeatletter\ifx\SetFigFont\undefined%
\gdef\SetFigFont#1#2#3#4#5{%
  \reset@font\fontsize{#1}{#2pt}%
  \fontfamily{#3}\fontseries{#4}\fontshape{#5}%
  \selectfont}%
\fi\endgroup%
{\renewcommand{\dashlinestretch}{30}
\begin{picture}(7894,4604)(0,-10)
\put(1981,1372){\makebox(0,0)[lb]{\smash{{{\SetFigFont{12}{14.4}{\familydefault}{\mddefault}{\updefault}$r-\lambda$}}}}}
\thicklines
\path(2880,517)(2882,517)(2887,518)
	(2896,520)(2909,523)(2927,526)
	(2949,531)(2974,536)(3003,542)
	(3033,548)(3064,554)(3095,561)
	(3125,567)(3153,573)(3180,579)
	(3204,585)(3227,591)(3247,596)
	(3267,601)(3285,607)(3305,613)
	(3324,620)(3343,627)(3361,634)
	(3379,641)(3396,648)(3413,655)
	(3430,663)(3447,670)(3463,678)
	(3480,685)(3497,693)(3514,701)
	(3531,709)(3548,717)(3565,725)
	(3583,733)(3600,742)(3617,751)
	(3635,760)(3652,769)(3670,779)
	(3687,788)(3705,797)(3722,807)
	(3739,817)(3757,826)(3774,836)
	(3791,846)(3809,856)(3826,866)
	(3844,877)(3861,888)(3879,899)
	(3897,910)(3915,922)(3931,933)
	(3948,945)(3966,958)(3986,972)
	(4006,987)(4028,1004)(4052,1022)
	(4076,1041)(4101,1061)(4127,1080)
	(4151,1100)(4175,1118)(4195,1134)
	(4213,1148)(4228,1159)(4238,1168)
	(4245,1173)(4249,1177)(4251,1178)
\put(4251,1178){\blacken\ellipse{60}{60}}
\put(4251,1178){\ellipse{60}{60}}
\path(1080,517)(7380,517)
\path(7140.000,457.000)(7380.000,517.000)(7140.000,577.000)
\path(2880,517)(2880,457)
\path(1470,1937)(1530,1937)
\dottedline{45}(1530,717)(3555,717)
\dottedline{45}(3555,717)(3555,517)
\dottedline{45}(4251,1178)(4251,517)
\path(3765,517)(1530,2797)
\path(1530,22)(1530,4567)
\path(1590.000,4327.000)(1530.000,4567.000)(1470.000,4327.000)
\dottedline{45}(1530,3899)(4912,517)
\put(1260,202){\makebox(0,0)[lb]{\smash{{{\SetFigFont{12}{14.4}{\familydefault}{\mddefault}{\updefault}$0$}}}}}
\put(7110,67){\makebox(0,0)[lb]{\smash{{{\SetFigFont{12}{14.4}{\familydefault}{\mddefault}{\updefault}$\lambda$}}}}}
\put(1170,2767){\makebox(0,0)[lb]{\smash{{{\SetFigFont{12}{14.4}{\familydefault}{\mddefault}{\updefault}$r$}}}}}
\put(675,1912){\makebox(0,0)[lb]{\smash{{{\SetFigFont{12}{14.4}{\familydefault}{\mddefault}{\updefault}$D(\rho||\sigma)$}}}}}
\put(5715,2317){\makebox(0,0)[lb]{\smash{{{\SetFigFont{12}{14.4}{\familydefault}{\mddefault}{\updefault}$\phi(\lambda)$}}}}}
\put(3465,157){\makebox(0,0)[lb]{\smash{{{\SetFigFont{12}{14.4}{\familydefault}{\mddefault}{\updefault}$\lambda^*$}}}}}
\put(4050,157){\makebox(0,0)[lb]{\smash{{{\SetFigFont{12}{14.4}{\familydefault}{\mddefault}{\updefault}$\psi'(1)$}}}}}
\put(0,697){\makebox(0,0)[lb]{\smash{{{\SetFigFont{12}{14.4}{\familydefault}{\mddefault}{\updefault}$\phi(\lambda^*)=r-\lambda^*$}}}}}
\put(2520,157){\makebox(0,0)[lb]{\smash{{{\SetFigFont{12}{14.4}{\familydefault}{\mddefault}{\updefault}$D(\rho||\sigma)$}}}}}
\put(4726,157){\makebox(0,0)[lb]{\smash{{{\SetFigFont{12}{14.4}{\familydefault}{\mddefault}{\updefault}$2\psi'(1)-\psi(1)$}}}}}
\put(91,3892){\makebox(0,0)[lb]{\smash{{{\SetFigFont{12}{14.4}{\familydefault}{\mddefault}{\updefault}$2\psi'(1)-\psi(1)$}}}}}
\path(4251,1178)(6079,2988)
\end{picture}
}
\end{center}
\caption{The graph of $\phi(\lambda)$}
\label{fig:phi}
\end{figure}
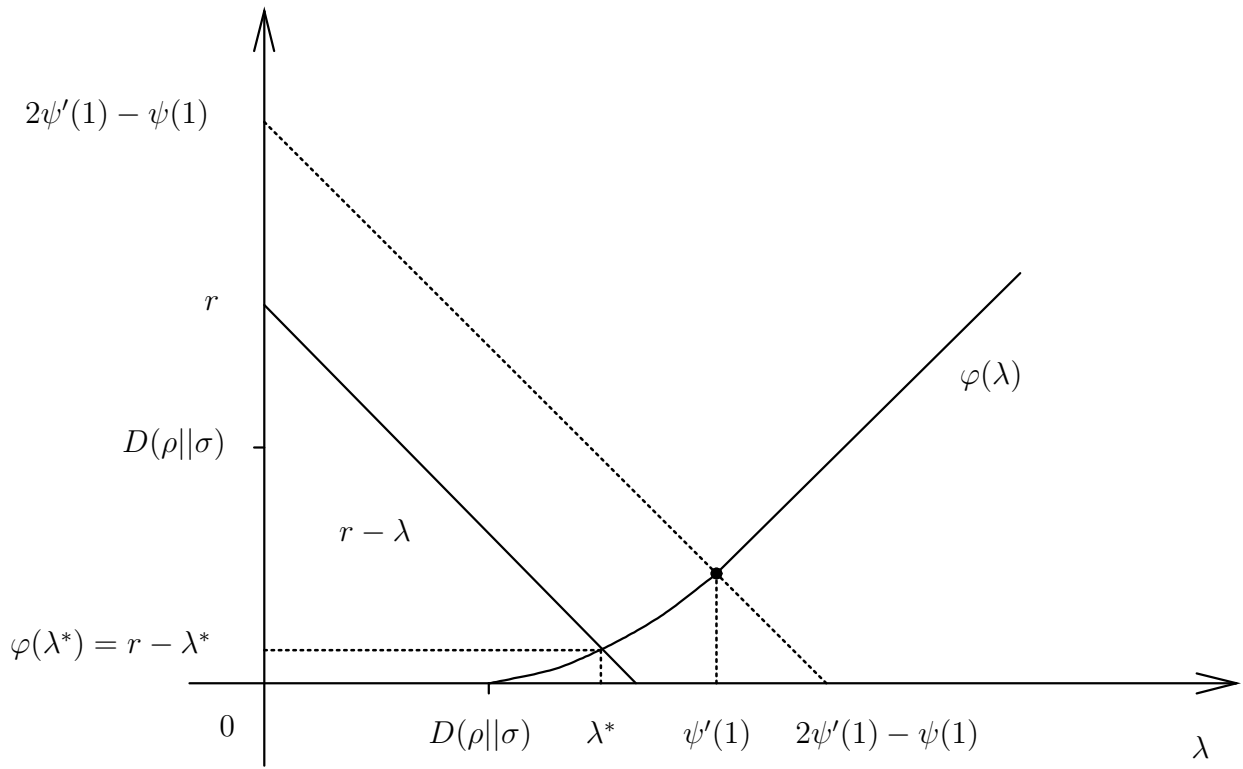

%%%%%%%%%%%%%%%%%%%%%%%%%%%%%%%%%%%%%%%%%%%%%%%%%%%%%%%%%%%%%%%%%%%%%%%%
\begin{theorem}
For any test $A_n$, if
\begin{eqnarray}
\limsup_{n\rightarrow\infty}\frac{1}{n}\log \beta_n(A_n) \le -r ,
\label{strong-converse-rate-if}
\end{eqnarray}
then
\begin{eqnarray}
\limsup_{n\rightarrow\infty}\frac{1}{n}\log (1-\alpha_n(A_n))
\le -\phi (\lambda^*) ,
\label{strong-converse-rate-then}
\end{eqnarray}
where $\lambda^*$ is a real number which
satisfies $\phi (\lambda^*) = r - \lambda^*$.
Moreover, $\phi(\lambda^*)$ is represented as
\begin{eqnarray}
\phi(\lambda^*) = \max_{0\le s \le 1}
\left\{
\frac{s}{1+s}r - \frac{1}{1+s} \psi(s)
\right\} .
\label{phi(lambda*)-max-representation}
\end{eqnarray}
\label{theorem:strong-converse-rate}
\end{theorem}
%%%%%%%%%%%%%%%%%%%%%%%%%%%%%%%%%%%%%%%%%%%%%%%%%%%%%%%%%%%%%%%%%%%%%%%%
\begin{proof}
For all $\delta>0$, there exists $n_0$ such that
\begin{eqnarray*}
\beta_n(A_n)\le e^{-n(r-\delta)}, \quad \forall n \ge n_0 ,
\end{eqnarray*}
from (\ref{strong-converse-rate-if}).
Putting $\lambda=\lambda^*$ in (\ref{fundamental}), we have
\begin{eqnarray*}
1-\alpha_n(A_n)\le e^{-n\phi(\lambda^*)}+e^{-n(r-\lambda^*-\delta)},
\quad \forall n \ge n_0 ,
\end{eqnarray*}
and hence,
\begin{eqnarray*}
\limsup_{n\rightarrow\infty}\frac{1}{n}\log (1-\alpha_n(A_n))
\le -\phi (\lambda^*) + \delta .
\end{eqnarray*}
Since $\delta>0$ is arbitrary,
(\ref{strong-converse-rate-then}) has been proved.

To show (\ref{phi(lambda*)-max-representation}),
suppose that
$\psi'(0)\le r\le 2\psi'(1)-\psi(1)$ firstly (see Fig. \ref{fig:phi}),
and $s^*$ attains the maximum in the equation
\begin{eqnarray*}
u(r)\Def \phi(\lambda^*) =
\max_{0\le s \le 1} \left\{
s\lambda^* - \psi(s)
\right\}
= r - \lambda^* .
\end{eqnarray*}
Then, taking (\ref{Legendre-trans}) into account,
$u(r)$ is represented parametrically as
\begin{eqnarray}
&& u(r) = s^* \psi'(s^*) -\psi(s^*) , \label{u(r):1} \\
\mbox{where,} && r = (s^*+1) \psi'(s^*) - \psi(s^*) . \label{u(r):2}
\end{eqnarray}
By using (\ref{u(r):2})
to eliminate $\psi'(s^*)$ from (\ref{u(r):1}),
we have
\begin{eqnarray*}
u(r) = \frac{s^*}{s^*+1}r -\frac{1}{s^*+1}\psi(s^*) .
\end{eqnarray*}
On the other hand, let
\begin{eqnarray*}
g(s) \Def \frac{s}{s+1}r -\frac{1}{s+1}\psi(s) ,
\end{eqnarray*}
then we have
\begin{eqnarray*}
g'(s)=\frac{1}{(s+1)^2} \left\{
r+\psi(s) - (1+s)\psi'(s)
\right\} .
\end{eqnarray*}
To examine the sign of $g'(s)$,
put $h(s)\Def r+\psi(s) - (1+s)\psi'(s)$, and
we have $h'(s)=-(1+s)\psi''(s) \le 0$, which indicates that
the sign of $g'(s)$ changes at most once. Therefore $g(s)$ takes its
maximum at $s=\hat{s}$ if and only if
\begin{eqnarray*}
r+\psi(\hat{s})-(1+\hat{s})\psi'(\hat{s}) =0 .
\end{eqnarray*}
This is nothing but the condition (\ref{u(r):2}),
and hence, we obtain $u(r)=\max_{0\le s \le 1} g(s)$.

In the other cases, it is clear that
\begin{eqnarray*}
\phi(\lambda^*) = \frac{1}{2}r - \frac{1}{2} \psi(1)
= g(1) = \max_{0\le s \le 1} g(s) ,
\quad \mbox{if}\quad r \ge 2\psi'(1)-\psi(1) ,
\end{eqnarray*}
and
\begin{eqnarray*}
\phi(\lambda^*) = 0
= g(0) = \max_{0\le s \le 1} g(s) ,
\quad \mbox{if}\quad r \le \psi'(0) .
\end{eqnarray*}
\end{proof}
%%%%%%%%%%%%%%%%%%%%%%%%%%%%%%%%%%%%%%%%%%%%%%%%%%%%%%%%%%%%%%%%%%%%%%%%
\SPACE

It should be noted that
(\ref{phi(lambda*)-max-representation}) corresponds
to the representation which Blahut \cite{Blahut} derived,
in the classical hypothesis testing
(i.e., when $\rho$ and $\sigma$ commute),
concerning the power exponent for $\alpha_n(A_n)$
when $r<D(\rho||\sigma)$.
We can easily see that if $r>D(\rho||\sigma)$ then $\phi(\lambda^*)>0$
(see Fig. \ref{fig:phi}). Hence, the following corollary holds.

%%%%%%%%%%%%%%%%%%%%%%%%%%%%%%%%%%%%%%%%%%%%%%%%%%%%%%%%%%%%%%%%%%%%%%%%
\begin{corollary}
For any test $A_n$, if
\begin{eqnarray}
\limsup_{n\rightarrow\infty} \frac{1}{n}\log \beta_n(A_n)
< -D(\rho||\sigma) ,
\label{strong-converse-if}
\end{eqnarray}
then $\alpha_n(A_n)$ goes to one exponentially.
\label{corollary:strong-converse}
\end{corollary}
%%%%%%%%%%%%%%%%%%%%%%%%%%%%%%%%%%%%%%%%%%%%%%%%%%%%%%%%%%%%%%%%%%%%%%%%

\section{Relation with the Classical Hypothesis Testing}

In this section, we discuss the relation between $\phi(\lambda^*)$ and
the power exponent derived by Han and Kobayashi
\cite{Han-Kobayashi}
in the classical hypothesis testing.

%
% classical hypothesis testing
%
Let $p$ and $q$ be probability distributions on a finite set $\X$, a
null hypothesis and an alternative hypothesis respectively.  And define
$\alpha_n(\A_n)\Def p^n(\A_n^c)$ and $\beta_n(\A_n)\Def q^n(\A_n)$,
where $p^n$ and $q^n$ are the {i.i.d.}~extensions of $p$ and $q$, and
$\A_n\subset \X^n$ is an acceptance region of $p^n$.  Blahut
\cite{Blahut} proved that if $\beta_n(\A_n) \le e^{-nr}\,(r>D(p||q))$
then $\alpha_n(\A_n)$ tends to one as $n\rightarrow\infty$ for all
$\A_n\subset\X^n$.  Although Blahut showed that $1-\alpha_n(A_n)$
converges at one in the polynomial order, Han and Kobayashi
\cite{Han-Kobayashi} proved a stronger result. 
Putting
\begin{eqnarray*}
\alpha_n^*(r) \Def
\min \{ \alpha_n(\A_n) \,|\, \A_n \subset \X^n,\,
        \beta_n(\A_n) \le e^{-nr} \} ,
\end{eqnarray*}
they derived the power
exponent for $1-\alpha_n^*(r)$, namely,
they proved
\begin{eqnarray}
\liminf_{n\rightarrow\infty} \frac{1}{n}\log (1-\alpha_n^*(r))
= -\tilde{u}(r) ,
\label{Han-Kobayashi-strong-converse-rate-equal}
\end{eqnarray}
where
\begin{eqnarray}
\tilde{u}(r) \Def
\min_{\hat{p}: D(\hat{p}||q)\le r}
\left\{ D(\hat{p}||p)+r-D(\hat{p}||q) \right\} .
\label{Han-Kobayashi-strong-converse-rate}
\end{eqnarray}

As remarked in Han and Kobayashi
\cite{Han-Kobayashi}, when $r > D(p||q)$ is not so large, the
minimum of (\ref{Han-Kobayashi-strong-converse-rate}) is attained with
equality, which we suppose here.
Applying the method used in Appendix,
(\ref{Han-Kobayashi-strong-converse-rate}) is rewritten as
\begin{eqnarray}
\tilde{u}(r) =
\max_{s\ge 0} \left\{
\frac{s}{1+s}r - \frac{1}{1+s} \log \sum_{j\in\X} p_j^{1+s}q_j^{-s}
\right\} .
\label{Han-Kobayashi-max-representation}
\end{eqnarray}
Moreover, when $r > D(p||q)$ is sufficiently small,
(\ref{Han-Kobayashi-max-representation}) yields
\begin{eqnarray*}
\tilde{u}(r) =
\max_{0 \le s \le 1} \left\{
\frac{s}{1+s}r - \frac{1}{1+s} \log \sum_{j\in\X} p_j^{1+s}q_j^{-s}
\right\} ,
\end{eqnarray*}
which corresponds to (\ref{phi(lambda*)-max-representation}).

It is interesting to observe that in the quantum case we have
\begin{eqnarray*}
\min_{\hat{\rho}: D(\hat{\rho}||\sigma)\le r}
\left\{ D(\hat{\rho}||\rho)+r-D(\hat{\rho}||\sigma) \right\}
=
\max_{s\ge 0} \left\{
\frac{s}{1+s}r - \frac{1}{1+s} \overline{\psi}(s)
\right\} ,
\end{eqnarray*}
where we put
\begin{eqnarray*}
\overline{\psi}(s) = \log \Tr e^{(1+s)\log\rho-s\log\sigma} .
\end{eqnarray*}
By the Golden-Thompson inequality
(see {\it e.g.} \cite{Bhatia}, p.261),
% Ohya-Petz p. 54 or 128
we can see that $\psi(s)\ge\overline{\psi}(s)$ and the equality holds
if and only if $\rho$ and $\sigma$ commute. Thus, we have
\begin{eqnarray}
\min_{\hat{\rho}: D(\hat{\rho}||\sigma)\le r}
\left\{ D(\hat{\rho}||\rho)+r-D(\hat{\rho}||\sigma) \right\}
\ge \phi(\lambda^*) .
\end{eqnarray}

\section{Concluding Remarks}

So far we have shown the fundamental inequality, and seen that
the quantum Stein's lemma and the strong converse
in the quantum hypothesis testing are obtained
as applications of the inequality.

In the classical hypothesis testing, $\tilde{u}(r)$ is shown to be the
optimal exponent. However, whether $\phi(\lambda^*)$ in the quantum
hypothesis testing is optimal or not is left open.

\appendix

We show (\ref{Han-Kobayashi-max-representation}) for readers'
convenience.
Here, we will derive (\ref{Han-Kobayashi-max-representation})
by the information geometrical method (see e.g., \cite{Amari}),
although (\ref{Han-Kobayashi-max-representation}) can be
shown by using the Lagrange multiplier method as given in \cite{Blahut}.

Suppose that $r>D(p||q)$ is not so large that
there exists a probability distribution of the form:
\begin{eqnarray*}
&&p(s)_j \Def e^{-\tipsi (s)} p^{1+s}_j q^{-s}_j ,
\quad (j\in\X,\,s > 0), \\
\mbox{where,} &&
\tipsi (s) \Def \log \sum_{j\in\X} p^{1+s}_j q^{-s}_j ,
\end{eqnarray*}
such that $D(p(s)||q)=r$.
Note that
\begin{eqnarray*}
\tipsi' (s) &=& E_{p(s)} \left[ \log \frac{p}{q} \right]
  \Def \eta(s) , \\
\tipsi'' (s) &=& E_{p(s)} \left[
  \left( \log \frac{p}{q} - \eta(s) \right)^2
\right] > 0 .
\end{eqnarray*}
Firstly, we will show that for all probability distribution $\hat{p}$
it holds that
\begin{eqnarray}
D(\hat{p}||q) = D(p(s)||q)
\Longrightarrow D(\hat{p}||p) \ge D(p(s)||p) .
\label{appendix:1}
\end{eqnarray}
To this end,
suppose that there exists a probability distribution $\hat{p}$
which satisfies
$D(\hat{p}||q) = D(p(s)||q)$ and
\begin{eqnarray*}
E_{p(s)} \left[ \log \frac{p}{q} \right]
< E_{\hat{p}} \left[ \log \frac{p}{q} \right] .
\end{eqnarray*}
Then, we have
\begin{eqnarray*}
D(\hat{p}||q) &=& D(\hat{p}||p(s))+D(p(s)||q)
  + \sum_{j\in\X} ( \hat{p}_j - p(s)_j )( \log p(s)_j -\log q_j ) \\
&=& D(\hat{p}||p(s))+D(p(s)||q)
  + \sum_{j\in\X} ( \hat{p}_j - p(s)_j )
    \left(
      (1+s) \log \frac{p_j}{q_j} - \tipsi(s)
    \right) \\
&=& D(\hat{p}||p(s))+D(p(s)||q)
    +(1+s) \left(
      E_{\hat{p}}\left[ \log \frac{p}{q} \right]
     -E_{p(s)}\left[ \log \frac{p}{q} \right]
    \right) \\
&>& D(p(s)||q) ,
\end{eqnarray*}
which contradicts with the assumption.
Therefore, for all probability distribution
with $D(\hat{p}||q) = D(p(s)||q)$, we have
\begin{eqnarray*}
E_{p(s)} \left[ \log \frac{p}{q} \right]
\ge E_{\hat{p}} \left[ \log \frac{p}{q} \right] ,
\end{eqnarray*}
and hence, there exists $t \le s$ such that
\begin{eqnarray}
E_{p(t)} \left[ \log \frac{p}{q} \right]
= E_{\hat{p}} \left[ \log \frac{p}{q} \right] ,
\label{appendix:2}
\end{eqnarray}
since $\eta(s)$ is continuous and monotone increasing.
Now, from (\ref{appendix:2}) and
the Pythagorean relation for the Kullback-Leibler divergence
we have
\begin{eqnarray}
D(\hat{p}||q) &=& D(\hat{p}||p(t)) + D(p(t)||q) \nonumber \\
&=& D(p(s)||q) .
\label{appendix:3}
\end{eqnarray}
Using the Pythagorean relation one more times
and from (\ref{appendix:3}),
(\ref{appendix:1}) is proved as follows:
\begin{eqnarray*}
D(\hat{p}||p) - D(p(s)||p)
&=& D(\hat{p}||p(t)) + D(p(t)||p) - D(p(s)||p) \\
&=& D(p(s)||q) - D(p(t)||q) +  D(p(t)||p) - D(p(s)||p) \\
&=& \left\{ D(p(s)||q) - D(p(s)||p) \right\}
    - \left\{ D(p(t)||q) -  D(p(t)||p) \right\} \\
&=& \eta(s)-\eta(t) \\
&\ge& 0 .
\end{eqnarray*}
Now, taking (\ref{appendix:1}) into account,
(\ref{Han-Kobayashi-strong-converse-rate})
is represented as
\begin{eqnarray}
\tilde{u}(r)
&=& \min_{s: D(p(s)||q)\le r}
\left\{ D(p(s)||p)+r-D(p(s)||q) \right\} \nonumber \\
&=& \min_{s: D(p(s)||q)\le r}
\left\{ r-\eta(s) \right\} .
\label{appendix:4}
\end{eqnarray}
Here, we can see that $d(s)\Def D(p(s)||q)$
is a monotone increasing function of $s$, which is verified by
$d'(s)=(1+s)\tipsi'' (s) >0$.
Thus, the minimum of (\ref{appendix:4}) is attained with equality,
and $\tilde{u}(r)$ is represented parametrically as
\begin{eqnarray*}
&& \tilde{u}(r) = D(p(s)||p) = s \,\eta(s) - \tipsi(s) , \\
\mbox{where,} && r = D(p(s)||q) = (1+s)\eta(s) -\tipsi(s) .
\end{eqnarray*}
This representation corresponds to (\ref{u(r):1}) (\ref{u(r):2})
and we obtain (\ref{Han-Kobayashi-max-representation})
by following the same procedure as
the proof of Theorem \ref{theorem:strong-converse-rate}.

\section*{Acknowledgment}

The authors wish to thank Prof.~Fumio~Hiai for his helpful comments.
They are also grateful to Prof.~Te~Sun~Han for his suggestion
for the strong converse in the quantum hypothesis testing.

\newpage

\newpage

\end{document}